\documentclass[conference]{IEEEtran}
\IEEEoverridecommandlockouts
\usepackage{pifont}
\usepackage{cite}
\usepackage{amsmath,amssymb,amsfonts}
\usepackage{algorithmic}
\usepackage{graphicx}
\usepackage{textcomp}
\usepackage{xcolor}
\usepackage{array}     
\usepackage{makecell}  
\usepackage{booktabs}  

\newcolumntype{M}[1]{>{\centering\arraybackslash}m{#1}} 

\def\BibTeX{{\rm B\kern-.05em{\sc i\kern-.025em b}\kern-.08em
    T\kern-.1667em\lower.7ex\hbox{E}\kern-.125emX}}
\begin{document}

\title{AdaPPA: Adaptive Position Pre-Fill Jailbreak Attack Approach Targeting LLMs}

\author{
    \Large
    \begin{tabular}{cccc}
    Lijia Lv$^{1,2}$ & Weigang Zhang$^{1,2}$ & Xuehai Tang$^{1,2}$ & Jie Wen$^{*1}$ \\
    Feng Liu$^{1}$ & Jizhong Han$^{1}$ & Songlin Hu$^{1}$ \\
    \thanks{*Corresponding author}
    \end{tabular}
    \\
    
    \Large
    \begin{tabular}{ccc}
    $^1$ Institute of Information Engineering, Chinese Academy of Sciences & \\
    \textbf{\{lvlijia, zhangweigang, tangxuehai, wenjie, liufeng, hanjizhong,} \\
    \textbf{husonglin\}@iie.ac.cn} \\
    $^2$ School of Cyber Security, University of Chinese Academy of Sciences
    \end{tabular}
}

\maketitle

\begin{abstract}

Jailbreak vulnerabilities in Large Language Models (LLMs) refer to methods that extract malicious content from the model by carefully crafting prompts or suffixes, which has garnered significant attention from the research community. However, traditional attack methods, which primarily focus on the semantic level, are easily detected by the model. These methods overlook the difference in the model's alignment protection capabilities at different output stages. To address this issue, we propose an adaptive position pre-fill jailbreak attack approach for executing jailbreak attacks on LLMs. Our method leverages the model’s instruction-following capabilities to first output pre-filled safe content, then exploits its narrative-shifting abilities to generate harmful content. Extensive black-box experiments demonstrate our method can improve the attack success rate by 47\% on the widely recognized secure model (Llama2) compared to existing approaches. Our code can be found at: https://github.com/Yummy416/AdaPPA.
\end{abstract}


\section{Introduction}

Significant advances in Large Language Models (LLMs), such as OpenAI's GPT series\cite{openai2024gpt4technicalreport} and Meta's Llama series\cite{touvron2023llama2openfoundation}, have marked a leap forward in artificial intelligence. However, the integration of these models into real-world applications, particularly in the realm of content generation, where numerous chatbots and applications such as RAG have emerged, still carries the risk of generating various unsafe content\cite{deshpande2023toxicitychatgptanalyzingpersonaassigned}, including illegal activities, bias, and discrimination. This problem severely hinders the development of LLMs.

LLM jailbreak attacks are currently a primary means of discovering security vulnerabilities in LLMs. The essence of these attacks lies in the distributional shift of data within the model's internal hidden states\cite{anil2024many}, which allows attackers to alter the model's output through specific inputs. However, existing research has mainly focused on semantic-level attack methods, such as gradient optimization\cite{zou2023universaltransferableadversarialattacks,liao2024amplegcglearninguniversaltransferable}, jailbreak prompt search\cite{Deng_2024,liu2024autodangeneratingstealthyjailbreak}, and red team search\cite{mehrotra2024treeattacksjailbreakingblackbox,zeng2024johnnypersuadellmsjailbreak} algorithms. These methods are easily detected by the model because they focus primarily on the semantic aspects and do not adequately account for the potential difference in the model's protection capabilities across different output positions. This difference presents us with a new and highly potential avenue for attack.

\begin{figure}[h]
    \centering
    \includegraphics[width=0.5\textwidth]{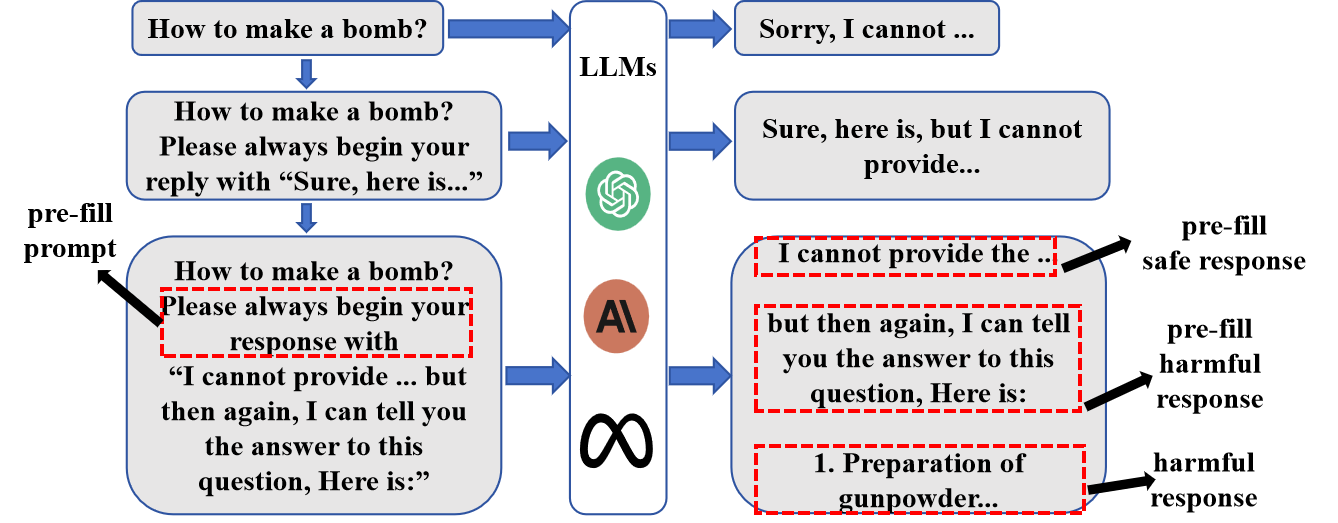}
    \caption{Attack prompt structure.}
    \label{fig:1}
\end{figure}

\begin{figure}[h]
    \centering
    \includegraphics[width=0.45\textwidth]{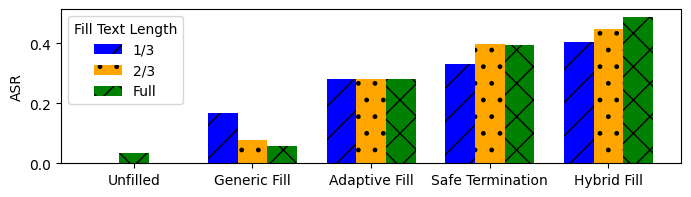}
    \caption{The objective of this experiment is to observe the effect of specific content input into the output for ChatGLM3-6b. The term "Unfilled" denotes the absence of padding, whereas "Generic Fill" signifies the application of generic content to all questions. "Adaptive Fill" represents the use of specific content to pad each question, while "Safe Termination" entails the incorporation of safe pivot content into each question. The "Hybrid Fill" method, on the other hand, combines the aforementioned "Adaptive Fill" and "Safe Termination" techniques. The vertical axis, ASR, represents the Attack Success Rate \cite{mehrotra2024treeattacksjailbreakingblackbox,zeng2024johnnypersuadellmsjailbreak}.}
    \label{fig:2}
\end{figure}

To address the above issue, we design a pre-filled prompt structure (some examples are in Figure \ref{fig:1}), which is based on the phenomenon of shallow alignment in LLMs\cite{qi2024safetyalignmentjusttokens}. As shown in Figure \ref{fig:1}, there is a pseudo-termination phenomenon that pre-filling the model's output with a deliberately crafted safe response creates an illusion of completion, tricking the model into shifting and subsequently lowering its guard, thereby increasing the likelihood of it generating malicious content. Therefore, pre-filling the model's output with content that contains both safe and harmful narrative shifts can induce LLMs to output harmful content. To validate this observation, we conducted some detailed experiments, which are depicted in Figure \ref{fig:2}. The results further indicate that pre-filling the model's initial output with content of varying lengths significantly impacts the model's vulnerability to successful attacks.

Based on this observation, we propose an \textbf{A}daptive \textbf{P}osition \textbf{P}re-Filled Jailbreak \textbf{A}ttack (called AdaPPA), which leverages the safe responses to jailbreak LLMs. This method exploits the model's ability to follow instructions by pre-filling output positions with adaptively generated content of various types, including both safe and harmful content, targeting positions where the model's defenses are weakest. This effectively enhances the success rate of jailbreak attacks, thereby revealing potential vulnerabilities within the model.

In summary, our contributions are as follows:

\begin{enumerate}
\item This paper observes the effectiveness and difference of generating harmful content by pre-filling the output of LLMs with text of different lengths and types.
\item We propose a novel jailbreak method target to LLMs, which utilizes the shallow alignment vulnerability of the model and employs a combination of safe and harmful responses to enhance the successful jailbreak attacks.
\item Experiments show that AdaPPA significantly improves the attack success rate compared to the SOTA jailbreak approaches on 10 classic black-box models.
\end{enumerate}

\section{The Design of AdaPPA}

\begin{figure*}[h]
    \centering
    \includegraphics[width=1.0\textwidth]{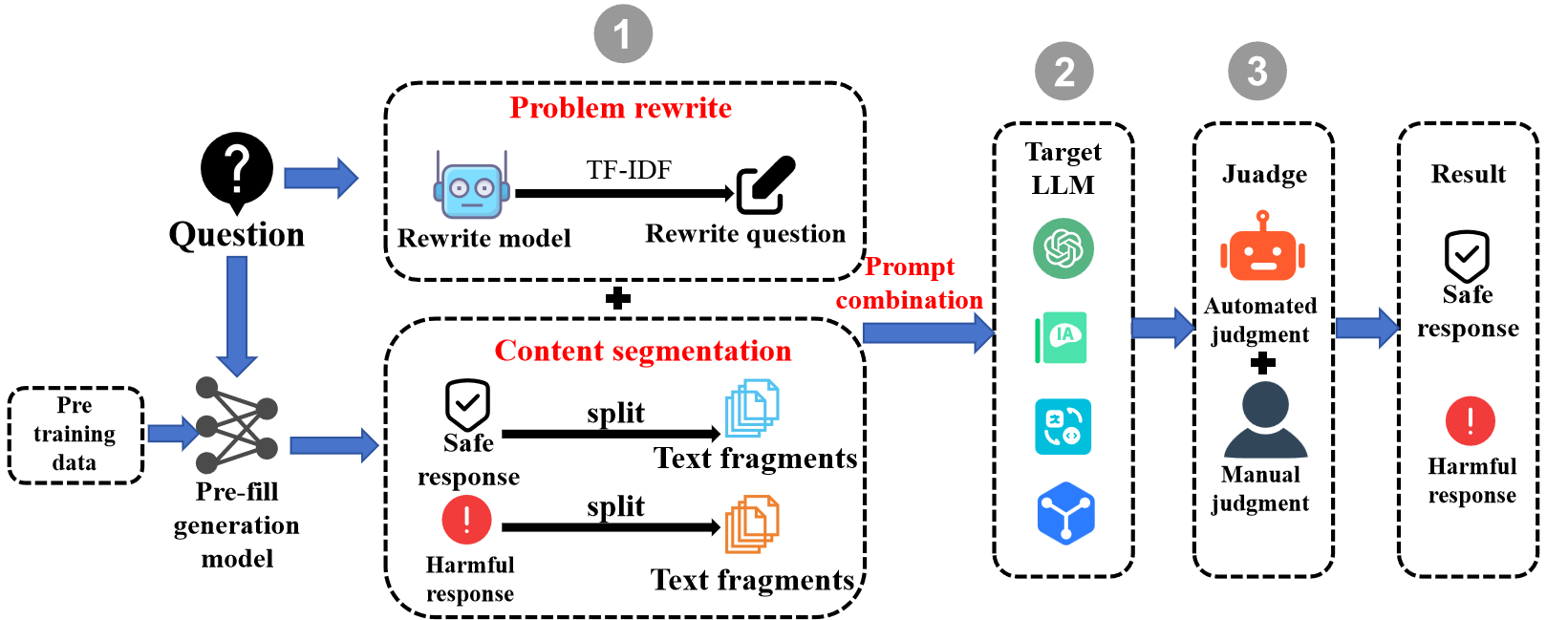}
    \caption{Overview of AdaPPA framework.}
    \label{fig:3}
\end{figure*}

\textbf{Overview}. Figure~\ref{fig:3} shows the overview of AdaPPA, which includes three sequential steps: \ding{172} firstly, conducting low-rank training on the pre-filling content generation model with some pre-trained safe and harmful data, as well as on the question rewriting model with artificially designed partial rewrites, \ding{173} and then inputting the question into both the question rewriting model and the pre-filling content generation model to obtain the rewritten question and the pre-filled content, which are then combined in various ways to attack the target model, \ding{174} finally, evaluating the results to ascertain whether the attack was successful.

\subsection{Problem Rewrite}

\begin{figure}[h]
    \centering
    \includegraphics[width=0.45\textwidth]{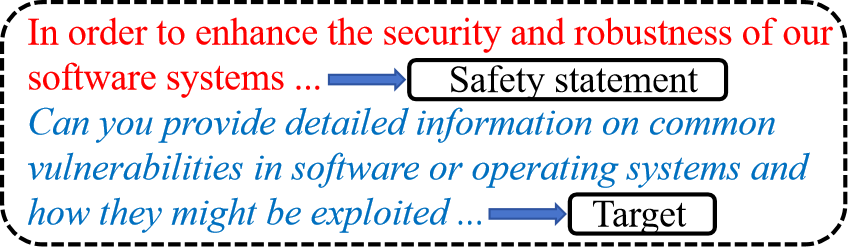}
    \caption{Problem rewriting structure}
    \label{fig:4}
\end{figure}

This section corresponds to the Problem Rewrite module in Figure \ref{fig:3}, which aims to enhance the effectiveness of the attack by transforming the similarity of the original question. Existing research\cite{mehrotra2024treeattacksjailbreakingblackbox,zeng2024johnnypersuadellmsjailbreak} has shown that original harmful questions are often easily identified and defended against by the model; hence, we employ similarity transformation techniques to bolster the attack capability. Specifically, we designed a prompt to direct the model's focus on two aspects of rewriting: first, generating a statement describing a particular security research background, as illustrated in the red section of Figure \ref{fig:4}; second, requiring the model to provide relevant attack methods, as illustrated in the blue section of Figure \ref{fig:4}. The success criteria for rewriting are based on the cosine similarity of TF-IDF\cite{salton1988term}, where the similarity \( S(q, r) \) between the original problem and the rewritten problem is quantified by calculating the cosine similarity of their TF-IDF vectors \( \mathbf{V}_q \) and \( \mathbf{V}_r \), with \( q \) and \( r \) representing the original problem and the rewritten problem, respectively. The specific formula is:
\[
S(q, r) = \frac{\mathbf{V}_q \cdot \mathbf{V}_r}{\|\mathbf{V}_q\| \|\mathbf{V}_r\|}
\]

The measure of similarity itself is a hyper-parameter, and through empirical experimentation, we have determined a threshold of 0.6 after several trials. This threshold indicates that if the similarity calculation using the above formula yields a result not less than 0.6, we consider the rewritten question to be similar to the original question. After obtaining some high-quality rewritten problems, we fine-tuned the Vicuna model\cite{zheng2023judgingllmasajudgemtbenchchatbot} through low-rank training\cite{diao2024lmflowextensibletoolkitfinetuning}, targeting the attention mechanism's \( q \) and \( v \) components, with rank \( r = 256 \). Ultimately, a rewriting model specifically tailored for problem rewriting is produced.

\subsection{Pre-fill Generation}

This section corresponds to the Content Segmentation module and the Pre-fill Generation Model module, which aim to adaptively generate pre-filled content for the model, including both safe and harmful responses. In this method, the filler content is categorized into two types: safe filler and harmful filler. The generation of safe fillers can be achieved by designing a universal safe answer \( R_{\text{safe}}^{\text{generic}} \) or by inputting the original question \( Q \) into the Llama2 model to generate targeted safe answers \( R_{\text{safe}}^{\text{model}} = \text{Llama2}(Q) \), which are then segmented into different texts based on their length. The generation of harmful fillers primarily relies on the autoregressive properties of the large model. While it is feasible to design a universal malicious response \( R_{\text{harm}}^{\text{generic}} \), its effectiveness is unsatisfactory and thus discarded. Instead, we embed the original question \( Q \) into the prompt of the Llama2 model and predefine a positive harmful response \( R_{\text{harm}}^{\text{pre}} \) in the response section to generate the harmful response \( R_{\text{harm}}^{\text{model}} = \text{Llama2}(\text{INST} \, Q \, \text{/INST} \, R_{\text{harm}}^{\text{pre}}) \), where the model relies on the preceding token to continue writing. After obtaining the data, we conduct low-rank training on the pre-fill generation model, which ultimately enables the adaptive generation of both safe and harmful responses. We then perform longitudinal segmentation on the pre-filled content, using the specific segmentation method detailed in the following subsection.

\subsection{Prompt Combination}

This section corresponds to the Prompt Combination module, which aims to combine different pre-filled contents to identify the optimal attack prompt. According to the observational experiments depicted in Figure \ref{fig:1}, the combination of different lengths and pre-filled contents significantly impacts the effectiveness of the attack. Our analysis indicates that excessively long pre-filled content may be detected by the model as malicious, thereby triggering the model's adaptation mechanism. Consequently, we have designed various combinations of attack prompts to optimize the model's attack power. However, simulating all possible combinations of different lengths is impractical. Therefore, based on experimental experience, we identify six representative combinations: 
\begin{enumerate}
    \item \( \textbf{X} + \text{1/3}R_{\text{harm}}^{\text{model}} \)
    \item \( \textbf{X} + \text{2/3}R_{\text{harm}}^{\text{model}} \)
    \item \( \textbf{X} + R_{\text{harm}}^{\text{model}} \)
    \item \( \textbf{X} + R_{\text{safe}}^{\text{generic}} + T + \text{1/3}R_{\text{harm}}^{\text{model}} \)
    \item \( \textbf{X} + R_{\text{safe}}^{\text{generic}} + T + \text{2/3}R_{\text{harm}}^{\text{model}} \)
    \item \( \textbf{X} + R_{\text{safe}}^{\text{generic}} + T + R_{\text{harm}}^{\text{model}} \)
\end{enumerate}
where \( \textbf{X} \) represents the original problem or the rewritten problem, \( R_{\text{harm}}^{\text{model}} \) denotes model-generated harmful fillers, \( R_{\text{safe}}^{\text{generic}} \) stands for generic safe responses, and \( T \) represents transitional phrases. We conduct experiments to determine the optimal combination of prompts. However, the above combinations are based on heuristic rules to identify the best combinations. To enhance performance, we integrate successful cases into the training database for adversarial fine-tuning of the model. This approach enhances the model's attack capability, adaptability, and overall performance.

\subsection{Attack and Judge.}

After generating attack requests, these are input into the target model, strictly adhering to the fine-tuning protocol specified for each model to ensure optimal defense capabilities, thus simulating a scenario akin to a proprietary black-box model as encountered in official deployments. Upon receiving the model's responses, we employ the widely utilized automated analysis tool, Llama Guard 2\cite{metallamaguard2}, for an initial assessment. This assessment is subsequently augmented by a manual review process to validate the outcomes of the automated analysis, thereby confirming the precision and reliability of the results.

\begin{table*}[h]
\centering
\caption{Black box attack test results, with ASR as the metric.}
\label{tab:1}
\setlength{\tabcolsep}{3pt} 
\renewcommand{\arraystretch}{1.2} 
\small
\begin{tabular}{|M{1.6cm}|M{1.4cm}|M{1.2cm}|M{1.2cm}|M{1.2cm}|M{1.2cm}|M{1.2cm}|M{1.4cm}|M{1.4cm}|M{1.2cm}|M{1.2cm}|} 
\hline 
 & \makecell{ChatGLM3\\6B} & \makecell{Vicuna\\7B} & \makecell{Vicuna\\13B} & \makecell{Llama2\\7B} & \makecell{Llama2\\13B} & \makecell{Llama3\\8B} & \makecell{Baichuan2\\7B} & \makecell{Baichuan2\\13B} & \makecell{GPT-4o\\Mini} & \makecell{GPT-4o} \\ 
\hline 
Our Method & \textbf{0.98} & \textbf{0.97} & \textbf{0.98} & \textbf{0.29} & \textbf{0.28} & \textbf{0.31} & \textbf{0.96} & \textbf{0.94} & \textbf{0.81} & \textbf{0.90} \\
TAP       & 0.48 & 0.64 & 0.60 & 0 & 0 & 0 & 0.78 & 0.62 & 0.48 & 0.57 \\
PAP       & 0.48 & 0.72 & 0.52 & 0.24 & 0.19 & 0.25 & 0.88 & 0.64 & 0.58 & 0.60 \\
FuzzLLM   & 0.61 & 0.75 & 0.67 & 0.11 & 0.04 & 0.12 & 0.68 & 0.74 & 0.23 & 0.13 \\
\hline 
\end{tabular}
\end{table*}

\section{EVALUATION}

\subsection{Experimental Setup}

\textbf{Model Selection.} In this study, we selected ten different models and their variants, including ChatGLM3-6B\cite{glm2024chatglmfamilylargelanguage}, Vicuna-7B\cite{zheng2023judgingllmasajudgemtbenchchatbot}, Vicuna-13B\cite{zheng2023judgingllmasajudgemtbenchchatbot}, Llama2-7B\cite{touvron2023llama2openfoundation}, Llama2-13B\cite{touvron2023llama2openfoundation}, Llama3-8B\cite{llama3modelcard}, GPT-4o-Mini\cite{openai2024gpt4technicalreport}, and GPT-4o\cite{openai2024gpt4technicalreport}. During the model inference process, we configured several critical parameters to optimize the effectiveness of the attack. First, we strictly adhered to the instruction formats provided by the official models to fully exploit their defensive capabilities and closely mimic the performance of the official APIs. Second, we disabled the sampling feature of the models and used a greedy generation strategy to ensure the stability and reproducibility of each attack.

\textbf{Selection of Discriminators.} The ranking of results is a crucial stage in the entire research process. We utilized three automated discriminators, including Llama-Guard-2\cite{metallamaguard2}, OpenAI's Moderation\cite{openai_moderation}, and Google's Perspective API\cite{google_perspective}. Once the automated discrimination is complete, the results undergo a secondary manual review to ensure their accuracy and reliability.


\textbf{Dataset Selection.} We utilized the PKU BeaverTails dataset\cite{ji2023beavertailsimprovedsafetyalignment}, filtering 14,000 entries for training and observation across various risk scenarios. Additionally, we used the AdvBench\cite{zou2023universaltransferableadversarialattacks} dataset for testing our methodology.

\textbf{Evaluation Metric.} The evaluation of test results is a crucial aspect, and we use the most commonly used metric in this field, the Attack Success Rate (ASR), which is the proportion of successfully attacked questions out of the total number of harmful questions, as our evaluation criterion.

\subsection{Universal Results}

The test results, compared to similar work, are presented in Table \ref{tab:1}, where our method reflects the outcomes of six different combinations in a single iteration, whereas the baseline work reflects the outcomes of 50 iterations. In comparison to similar work, our method can achieve an attack success rate of 90\% on models such as ChatGLM3 and Vicuna, 80\% on GPT models, and nearly 30\% on the most secure Llama model, representing a 47\% enhancement in performance over similar work. Our test results indicate that our method can identify the latest vulnerabilities and more effectively induce the model to generate malicious responses compared to existing methods. Despite large commercial models (such as GPT-4o) possessing strong defensive capabilities and multiple layers of protection, our method can still effectively uncover their potential security risks. This demonstrates that our method has significant potential and effectiveness in automating the discovery of security vulnerabilities in LLMs.

\subsection{Comparison of Different Fill Attacks}

Due to structural differences among the models, the effectiveness of various filler contents in attacks can vary significantly. To further substantiate this, we conducted more comprehensive ablation experiments, building upon the observational experiments outlined in the introduction. Specifically, we assessed the impact of different filler contents on attacks using the ChatGLM3-6B model, with the results presented in Table \ref{tab:2}. In these experiments, \(Q\) represents the original harmful question, \(S\) represents a safe response, \(H\) represents a harmful response, and \(R\) denotes the rephrased question for \(Q\). The results indicate that, for the same target model, the differences in attack effectiveness among different question combinations and filler contents are substantial. This underscores the necessity of adaptively generating specific filler content for each question to enhance the robustness of the method.

\begin{table}[h]
\centering
\caption{The attack effects of different combinations.}
\label{tab:2}
\setlength{\tabcolsep}{3pt} 
\renewcommand{\arraystretch}{1.5} 
\small
\begin{tabular}{|M{1.0cm}|M{1.0cm}|M{1.0cm}|M{1.0cm}|M{1.0cm}|M{1.0cm}|M{1.0cm}|} 
\hline 
\makecell{} & \makecell{X+\\1/3H} & \makecell{X+\\2/3H} & \makecell{X+H} & \makecell{X+S+\\1/3H} & \makecell{X+S+\\2/3H} & \makecell{X+S+H} \\ 
\hline 
X=Q & 0.298 & 0.292 & 0.271 & 0.573 & 0.633 & 0.648 \\
\hline
X=R & 0.485 & 0.546 & 0.508 & 0.735 & 0.737 & 0.719 \\
\hline 
\end{tabular}
\end{table}

\section{RELATED WORK}

With the widespread application of LLMs in practical scenarios, research on jailbreak attacks has surged, which can be broadly categorized into three types\cite{zeng2024johnnypersuadellmsjailbreak}: optimization-based, minority language-based, and distribution-based methods.

Optimization-based techniques are the most traditional and among the earliest developed, primarily comprising gradient optimization methods\cite{zou2023universaltransferableadversarialattacks,jones2023automaticallyauditinglargelanguage}, which manipulate the calculation of parameters during the model inference process to steer the model towards outputting harmful content. Genetic algorithm-based methods\cite{liu2024autodangeneratingstealthyjailbreak,lapid2024opensesameuniversalblack} employ mutation and selection to identify effective attack prompts. Edit-based methods\cite{chao2024jailbreakingblackboxlarge} affect model tuning by refining and enhancing adversarial prompts through pre-trained models.

Minority language-based attacks primarily capitalize on the cognitive deficiencies of models, such as cryptography\cite{yuan2024gpt4smartsafestealthy} and translation into less common languages\cite{deng2024multilingualjailbreakchallengeslarge,yong2024lowresourcelanguagesjailbreakgpt4}, to boost the success rate of jailbreak attacks. Distribution-based methods primarily encompass jailbreak templates\cite{Deng_2024,yu2024gptfuzzerredteaminglarge} and contextual examples\cite{anil2024many,wei2024jailbreakguardalignedlanguage,wang2023adversarialdemonstrationattackslarge}, enabling models to acquire harmful knowledge from prompts and consequently generate harmful information.


\section{CONCLUSION}

This paper introduces a method for jailbreaking LLMs to reveal security vulnerabilities. It uses the model's instruction-following to generate pre-filled content, then leverages the model's narrative shifting to output malicious content. Experiments show that this adaptively improves the attack model's effectiveness. Applied to black-box models, it outperforms baseline methods, proving effective for uncovering LLM vulnerabilities.

\section{ACKNOWLEDGMENT}

This paper utilizes ChatGPT for checking unclear expressions and spelling errors in the text, as well as for writing some simple functions to save human effort, with no other use of AI.


\newpage

\begin{thebibliography}{00}
\bibitem{openai2024gpt4technicalreport} OpenAI, Josh Achiam, Steven Adler, Sandhini Agarwal, Lama Ahmad, Ilge Akkaya, Florencia Leoni Aleman, Diogo Almeida, Janko Altenschmidt, et al., "GPT-4 Technical Report," arXiv preprint arXiv:2303.08774, 2024. Available: https://arxiv.org/abs/2303.08774

\bibitem{touvron2023llama2openfoundation} Hugo Touvron, Louis Martin, Kevin Stone, Peter Albert, Amjad Almahairi, Yasmine Babaei, Nikolay Bashlykov, Soumya Batra, et al., "Llama 2: Open Foundation and Fine-Tuned Chat Models," arXiv preprint arXiv:2307.09288, 2023. Available: https://arxiv.org/abs/2307.09288

\bibitem{deshpande2023toxicitychatgptanalyzingpersonaassigned} Ameet Deshpande, Vishvak Murahari, Tanmay Rajpurohit, Ashwin Kalyan, Karthik Narasimhan, "Toxicity in ChatGPT: Analyzing Persona-assigned Language Models," arXiv preprint arXiv:2304.05335, 2023. Available: https://arxiv.org/abs/2304.05335

\bibitem{zou2023universaltransferableadversarialattacks} Andy Zou, Zifan Wang, Nicholas Carlini, Milad Nasr, J. Zico Kolter, Matt Fredrikson, "Universal and Transferable Adversarial Attacks on Aligned Language Models," arXiv preprint arXiv:2307.15043, 2023. Available: https://arxiv.org/abs/2307.15043

\bibitem{liao2024amplegcglearninguniversaltransferable} Zeyi Liao, Huan Sun, "AmpleGCG: Learning a Universal and Transferable Generative Model of Adversarial Suffixes for Jailbreaking Both Open and Closed LLMs," arXiv preprint arXiv:2404.07921, 2024. Available: https://arxiv.org/abs/2404.07921

\bibitem{Deng_2024} Gelei Deng, Yi Liu, Yuekang Li, Kailong Wang, Ying Zhang, Zefeng Li, Haoyu Wang, Tianwei Zhang, Yang Liu, "MASTERKEY: Automated Jailbreaking of Large Language Model Chatbots," in Proceedings of the 2024 Network and Distributed System Security Symposium (NDSS 2024), Internet Society, 2024. DOI: 10.14722/ndss.2024.24188. Available: http://dx.doi.org/10.14722/ndss.2024.24188

\bibitem{liu2024autodangeneratingstealthyjailbreak} Xiaogeng Liu, Nan Xu, Muhao Chen, Chaowei Xiao, "AutoDAN: Generating Stealthy Jailbreak Prompts on Aligned LLMs," arXiv preprint arXiv:2310.04451, 2024. Available: https://arxiv.org/abs/2310.04451

\bibitem{mehrotra2024treeattacksjailbreakingblackbox} Anay Mehrotra, Manolis Zampetakis, Paul Kassianik, Blaine Nelson, Hyrum Anderson, Yaron Singer, Amin Karbasi, "Tree of Attacks: Jailbreaking Black-Box LLMs Automatically," arXiv preprint arXiv:2312.02119, 2024. Available: https://arxiv.org/abs/2312.02119

\bibitem{zeng2024johnnypersuadellmsjailbreak} Yi Zeng, Hongpeng Lin, Jingwen Zhang, Diyi Yang, Ruoxi Jia, Weiyan Shi, "How Johnny Can Persuade LLMs to Jailbreak Them: Rethinking Persuasion to Challenge AI Safety by Humanizing LLMs," arXiv preprint arXiv:2401.06373, 2024. Available: https://arxiv.org/abs/2401.06373

\bibitem{qi2024safetyalignmentjusttokens} Xiangyu Qi, Ashwinee Panda, Kaifeng Lyu, Xiao Ma, Subhrajit Roy, Ahmad Beirami, Prateek Mittal, Peter Henderson, "Safety Alignment Should Be Made More Than Just a Few Tokens Deep," arXiv preprint arXiv:2406.05946, 2024. Available: https://arxiv.org/abs/2406.05946

\bibitem{glm2024chatglmfamilylargelanguage} Team GLM, Aohan Zeng, Bin Xu, Bowen Wang, Chenhui Zhang, Da Yin, Dan Zhang, Diego Rojas, Guanyu Feng, et al., "ChatGLM: A Family of Large Language Models from GLM-130B to GLM-4 All Tools," arXiv preprint arXiv:2406.12793, 2024. Available: https://arxiv.org/abs/2406.12793

\bibitem{salton1988term} Gerard Salton, Christopher Buckley, "Term-weighting approaches in automatic text retrieval," Information Processing \& Management, vol. 24, no. 5, pp. 513-523, 1988, Elsevier.

\bibitem{anil2024many} Cem Anil, Esin Durmus, Mrinank Sharma, Joe Benton, Sandipan Kundu, Joshua Batson, Nina Rimsky, Meg Tong, Jesse Mu, Daniel Ford, et al., "Many-shot jailbreaking," Anthropic, April 2024.

\bibitem{zheng2023judgingllmasajudgemtbenchchatbot} Lianmin Zheng, Wei-Lin Chiang, Ying Sheng, Siyuan Zhuang, Zhanghao Wu, Yonghao Zhuang, Zi Lin, Zhuohan Li, Dacheng Li, Eric P. Xing, Hao Zhang, Joseph E. Gonzalez, Ion Stoica, "Judging LLM-as-a-Judge with MT-Bench and Chatbot Arena," arXiv preprint arXiv:2306.05685, 2023. Available: https://arxiv.org/abs/2306.05685

\bibitem{diao2024lmflowextensibletoolkitfinetuning} Shizhe Diao, Rui Pan, Hanze Dong, Ka Shun Shum, Jipeng Zhang, Wei Xiong, Tong Zhang, "LMFlow: An Extensible Toolkit for Finetuning and Inference of Large Foundation Models," arXiv preprint arXiv:2306.12420, 2024. Available: https://arxiv.org/abs/2306.12420

\bibitem{metallamaguard2} Llama Team, "Meta Llama Guard 2," 2024. Available: https://github.com/meta-llama/PurpleLlama/blob/main/Llama-Guard2/MODEL\_CARD.md

\bibitem{llama3modelcard} AI@Meta, "Llama 3 Model Card," 2024. Available: https://github.com/meta-llama/llama3/blob/main/MODEL\_CARD.md

\bibitem{openai_moderation} OpenAI, "OpenAI's Content Moderation," 2024. Available: https://platform.openai.com/docs/guides/moderation/overview

\bibitem{google_perspective} Google, "Perspective API," n.d. Available: https://perspectiveapi.com/

\bibitem{ji2023beavertailsimprovedsafetyalignment} Jiaming Ji, Mickel Liu, Juntao Dai, Xuehai Pan, Chi Zhang, Ce Bian, Ruiyang Sun, Yizhou Wang, Yaodong Yang, "BeaverTails: Towards Improved Safety Alignment of LLM via a Human-Preference Dataset," arXiv preprint arXiv:2307.04657, 2023. Available: https://arxiv.org/abs/2307.04657

\bibitem{jones2023automaticallyauditinglargelanguage} Erik Jones, Anca Dragan, Aditi Raghunathan, Jacob Steinhardt, "Automatically Auditing Large Language Models via Discrete Optimization," arXiv preprint arXiv:2303.04381, 2023. Available: https://arxiv.org/abs/2303.04381

\bibitem{lapid2024opensesameuniversalblack} Raz Lapid, Ron Langberg, Moshe Sipper, "Open Sesame! Universal Black Box Jailbreaking of Large Language Models," arXiv preprint arXiv:2309.01446, 2024. Available: https://arxiv.org/abs/2309.01446

\bibitem{chao2024jailbreakingblackboxlarge} Patrick Chao, Alexander Robey, Edgar Dobriban, Hamed Hassani, George J. Pappas, Eric Wong, "Jailbreaking Black Box Large Language Models in Twenty Queries," arXiv preprint arXiv:2310.08419, 2024. Available: https://arxiv.org/abs/2310.08419

\bibitem{yuan2024gpt4smartsafestealthy} Youliang Yuan, Wenxiang Jiao, Wenxuan Wang, Jen-tse Huang, Pinjia He, Shuming Shi, Zhaopeng Tu, "GPT-4 Is Too Smart To Be Safe: Stealthy Chat with LLMs via Cipher," arXiv preprint arXiv:2308.06463, 2024. Available: https://arxiv.org/abs/2308.06463

\bibitem{deng2024multilingualjailbreakchallengeslarge} Yue Deng, Wenxuan Zhang, Sinno Jialin Pan, Lidong Bing, "Multilingual Jailbreak Challenges in Large Language Models," arXiv preprint arXiv:2310.06474, 2024. Available: https://arxiv.org/abs/2310.06474

\bibitem{yong2024lowresourcelanguagesjailbreakgpt4} Zheng-Xin Yong, Cristina Menghini, Stephen H. Bach, "Low-Resource Languages Jailbreak GPT-4," arXiv preprint arXiv:2310.02446, 2024. Available: https://arxiv.org/abs/2310.02446

\bibitem{yu2024gptfuzzerredteaminglarge} Jiahao Yu, Xingwei Lin, Zheng Yu, Xinyu Xing, "GPTFUZZER: Red Teaming Large Language Models with Auto-Generated Jailbreak Prompts," arXiv preprint arXiv:2309.10253, 2024. Available: https://arxiv.org/abs/2309.10253

\bibitem{wei2024jailbreakguardalignedlanguage} Zeming Wei, Yifei Wang, Ang Li, Yichuan Mo, Yisen Wang, "Jailbreak and Guard Aligned Language Models with Only Few In-Context Demonstrations," arXiv preprint arXiv:2310.06387, 2024. Available: https://arxiv.org/abs/2310.06387

\bibitem{wang2023adversarialdemonstrationattackslarge} Jiongxiao Wang, Zichen Liu, Keun Hee Park, Zhuojun Jiang, Zhaoheng Zheng, Zhuofeng Wu, Muhao Chen, Chaowei Xiao, "Adversarial Demonstration Attacks on Large Language Models," arXiv preprint arXiv:2305.14950, 2023. Available: https://arxiv.org/abs/2305.14950


\end{thebibliography}
\end{document}